\documentclass[pre,twocolumn,superscriptaddress,aps,notitlepage,10pt]{revtex4-2}
\usepackage{natbib}
\usepackage{times}
\usepackage{amssymb}
\usepackage{amsmath}
\usepackage{bm}
\usepackage{color}
\usepackage[colorlinks=true,urlcolor=blue,breaklinks,citecolor=blue]{hyperref}
\usepackage{graphicx}
\usepackage{braket}
\usepackage{subfigure}
\usepackage{xcolor}
\usepackage[version=4]{mhchem}
\usepackage[utf8]{inputenc}
\usepackage{siunitx}

\begin{document}

\title{OnionVQE Optimization Strategy for Ground State Preparation on NISQ Devices}

\author{Katerina Gratsea}
\email[]{gratsea.katerina@gmail.com}
\affiliation{Robert Bosch LLC, Research and Technology Center, Sunnyvale, CA 94085, USA}
\affiliation{ICFO-Institut  de  Ciencies  Fotoniques,  The  Barcelona  Institute  of Science  and  Technology, 08860  Castelldefels  (Barcelona),  Spain}
\author{Johannes Selisko}
\affiliation{Corporate Sector Research and Advance Engineering, Robert Bosch GmbH, Robert-Bosch-Campus 1, D-71272 Renningen, Germany}
\author{Maximilian Amsler}
\affiliation{Corporate Sector Research and Advance Engineering, Robert Bosch GmbH, Robert-Bosch-Campus 1, D-71272 Renningen, Germany}
\author{Christopher Wever}
\affiliation{Corporate Sector Research and Advance Engineering, Robert Bosch GmbH, Robert-Bosch-Campus 1, D-71272 Renningen, Germany}
\author{Thomas Eckl}
\affiliation{Corporate Sector Research and Advance Engineering, Robert Bosch GmbH, Robert-Bosch-Campus 1, D-71272 Renningen, Germany}
\author{Georgy Samsonidze}
\email[]{georgy.samsonidze@us.bosch.com}
\affiliation{Robert Bosch LLC, Research and Technology Center, Sunnyvale, CA 94085, USA}

\begin{abstract}

\textbf{Abstract} The Variational Quantum Eigensolver (VQE) is one of the most promising and widely used algorithms for exploiting the capabilities of current Noisy Intermediate-Scale Quantum (NISQ) devices. However, VQE algorithms suffer from a plethora of issues, such as barren plateaus, local minima, quantum hardware noise, and limited qubit connectivity, thus posing challenges for their successful deployment on hardware and simulators. In this work, we propose a VQE optimization strategy that builds upon recent advances in the literature, and exhibits very shallow circuit depths when applied to the specific system of interest, namely a model Hamiltonian representing a cuprate superconductor. These features make our approach a favorable candidate for generating good ground state approximations on current NISQ devices. Our findings illustrate the potential of VQE algorithmic development for leveraging the full capabilities of NISQ devices.


\end{abstract}

\maketitle

\section{Introduction}
\label{sec:Introduction}

In recent years, various flavors of the variational quantum eigensolver (VQE)~\cite{VQE_review, McClean_2016, Peruzzo_2014, HardEff, Economou} have been developed and extensively studied to serve as a hybrid quantum-classical algorithm on noisy intermediate-scale quantum (NISQ) devices. Despite the numerous variations, VQE algorithms face specific challenges, such as the optimal selection of an initial state~\cite{LA_initial}, identifying the most suitable ansatz~\cite{Zapata_circuits}, and experiencing slow convergence due to the presence of barren plateaus arising from the ill-conditioned optimization problem~\cite{Identity_Block_paper}. And even though VQE was initially designed to leverage the capabilities of both classical and quantum hardware~\cite{Peruzzo_2014, McClean_2016}, its performance on NISQ devices has thus far fallen short of noiseless simulations, necessitating the use of error mitigation techniques~\cite{H2_benchmark, HardEff, Google_overlaps}.

A plethora of recent studies have focused on either one of the problems individually that arise in VQE algorithms and have proposed various approaches to potentially overcome them~\cite{Identity, LA_initial, Identity_Block_paper, barren, wiersema2020identity}. Nevertheless, algorithmic development of VQE remains challenging and alternative methods have been proposed for ground state energy estimation, such as imaginary time evolution and robust amplitude estimation~\cite{alter1, alter2, alter3}. The recent work in Ref.~\cite{acc_critera} motivated the use of VQE algorithms to generate ground state approximations instead of focusing on the energy estimation. A few studies have already used VQE algorithms to generate ground state approximations on NISQ devices~\cite{100qubit_VQE, HF_Google, generative_ground_states, OIST, overlapadaptvqe, gratsea2024comparing}.

VQE implementations on current NISQ devices remain difficult due to the effect of noise from the quantum devices~\cite{HardEff, H2_benchmark}, but also due to challenges in algorithmic development, i.e., choice of the ansatz~\cite{Zapata_circuits} and presence of barren plateaus ~\cite{McClean_barren, larocca2024review}. To overcome these issues, error mitigation techniques have been applied on the hardware experiments. Additionally, the performance of the VQE algorithm has been enhanced through classically pre-optimizing the initial values for the parameterized gates~\cite{Identity, HF_Google, 100qubit_VQE}.

Despite recent advancements, the research community continues to seek a VQE algorithm capable of harnessing the potential of current NISQ devices. The critical questions are whether it is possible to pre-determine optimal initial values for parameterized gates, which can mitigate challenges associated with barren plateaus, and whether we can achieve a good ground state approximation, whose overlap with the exact ground state is sufficient for the task at hand~\cite{acc_critera, Andrew_2023}.

To this end, we propose a VQE optimization strategy that builds upon other recent works~\cite{Identity_Block_paper, HardEff, LA_initial}. Specifically, we start the quantum circuit with an entangled Hartree-Fock (HF) state and identify a hardware-friendly ansatz in the spirit of the hardware-efficient ansatz~\cite{HardEff} that results in a shallower circuit depth. Moreover, we propose a simple technique that initializes the parameterized unitary as the identity operator by leveraging the structure of the circuit we use. This is achieved by exploiting the linear CZ entangling map and initializing the parameterized gates in a layer-wise fashion to produce the identity operator, a method we will refer to as the onion-initialization. As discussed in Grant \textit{et al.}~\cite{Identity_Block_paper}, initializing the parameterized unitary close to the identity operator could potentially help with the barren plateau issue. We refer to the proposed optimization strategy as \textit{OnionVQE} to reflect both the onion-initialization, but also the combination of the different mechanisms discussed above (the choice of initial state, variational ansatz, and identity initialization) that play a role in its performance.

The OnionVQE optimization strategy suggests how to initialize the parameterized gates within specific circuit structures and uses a small number of operations with shallow circuit depths for our particular systems. This approach makes OnionVQE a favorable candidate for implementations on current NISQ devices, where minimizing the circuit depth and number of operations is of utmost importance. We explore the performance of the OnionVQE optimization strategy with an increasing system size up to $10$ qubits and under the effect of simulated noise from one of IBM's quantum devices.

We demonstrate the utility of OnionVQE by addressing a Hamiltionian derived from a real material system, \ce{CaCuO2}, which is one of the parent compounds for cuprate superconductors and exhibits properties stemming from strongly correlated electron-electron interactions~\cite{PhysRevB.39.9122}. We map the correlated $d_{x^2-y^2}$ state of \ce{CaCuO2} to an effective Hubbard Hamiltonian, and solve the associated Anderson impurity model (AIM) with dynamical mean field theory (DMFT) at varying number of bath sites (see Appendix~\ref{App:Hamiltonian} and Appendix~\ref{App:QHamiltonian})~\cite{selisko-24-dmft}. The bath in the AIM represents different energy scales, physical environments, or degrees of freedom in the system. To map the AIM to a quantum device, we discretize the continuum of the bath with a finite number of bath levels. In this study, we use AIM with $ \{ 1,2,3,4 \} $ bath sites, which map to $ \{ 4,6,8,10 \} $ qubit systems, respectively. We emphasize that the performance of the OnionVQE optimization strategy is independent of the studied system, i.e., it does not exploit specific symmetries of the associated Hamiltonians.

This paper is organized as follows. In Sec.~\ref{sec:VQE_opt_strategy}, we present the VQE optimization strategy along with the underlying components for its successful implementation. In Sec.~\ref{sec:Results_ibmFakeDevices}, we explore the performance of the OnionVQE algorithm on IBM's FakeMumbai. Finally, we present the conclusions and outlook in Sec.~\ref{sec:Conclusions}.

\section{Methodology}
\label{sec:VQE_opt_strategy}

In this section, we elaborate in more detail on the VQE optimization strategy employed to get good ground state approximations, i.e., with fidelity values above $0.94$ for all the systems we investigated. As already emphasized in the introduction, if our goal is to run the hybrid algorithm on current NISQ devices it is of utmost importance to reduce the circuit depth and number of entangling gates. To achieve this goal, we integrate three key components into our VQE optimization strategy that together produce good ground state approximations with shallow quantum circuits and a very small number of entangling gates for our Hamiltonians.

The three components of the VQE optimization strategy are:
\begin{itemize}
    \item Mean-field initial state: initialization of the circuit to a classically inspired state that is not a product state (see Sec.~\ref{subsec:mf}),
    \item Shallow parameterized quantum circuit structures (see Sec.~\ref{subsec:shallow-circuit}),
    \item Initialization of the parameterized unitary to the identity operator (see Sec.~\ref{subsec:IdBlock}).
\end{itemize}

The impacts of all the aforementioned components on VQE algorithms have been independently proposed and explored in various studies~\cite{LA_initial, unified, Identity_Block_paper}. In this paper, we show that by combining these components, we can circumvent the problem of barren plateaus related to the initialization of the parameterized part of the circuit for our particular systems, namely an Anderson impurity model (see Appendix~\ref{App:Hamiltonian}). Moreover, we have shallow circuit depths suitable for implementation on current NISQ devices.

\subsection{Mean-field initial state}
\label{subsec:mf}

Before we start optimizing the parameterized part of a quantum circuit, we need to decide how our qubits will be initialized, e.g., the zero state ($|00..00 \rangle $) or a classically inspired solution, such as a mean-field state. The mean-field Hartree-Fock (HF) state has extensively been used as an initial state in different VQE algorithms~\cite{SPA, VQE_review}. Once the Hamiltonian of the studied system is transformed to the molecular orbital (MO) basis, the HF state becomes a trivially preparable product state, such as $ | 000000111111 \rangle $ in the work of Arute \textit{et al.}~\cite{HF_Google}. To a certain extent this is equivalent to the zero state, in the sense that the initial state is simply a product state.

In our case, the Hamiltonian is more compactly expressed in the atomic orbital (AO) basis than in the molecular orbital (MO) basis. Specifically, the number of Pauli strings increases more than tenfold when transforming the Hamiltonian from the AO to MO basis (see Fig.~\ref{Afig:Pauli_strings} in Appendix~\ref{App:QHamiltonian}). Therefore, we opt to work in the AO basis. As a result, the HF state is not a product state; it already entangles the qubits (see Fig.~\ref{Afig:mf_4qubit} in Appendix~\ref{App:HF_State}). Notably, the choice of the initial state affects the trainability of a shallow-depth hardware-efficient type ansatz~\cite{LA_initial, unified}. As discussed in the aforementioned references, a little prior knowledge of the initial state could help avoid the effect of barren plateaus in VQE optimization, and moreover, an entangled initial state could also help with trainability issues~\cite{LA_initial, unified, Identity}.

The efficient decomposition of a given state to a quantum circuit is not a trivial task and a different set of gates could be used that are native to the underlying hardware platforms. In this paper, we focus on IBM devices and use the pre-defined decomposition function in Qiskit~\cite{jiang2018slaterdeterminant, qiskit2024}. The number of gates that constitute the HF circuit for the 4, 6, 8, and 10 qubit cases are shown in Fig.~\ref{Afig:mf_decomposition} in Appendix~\ref{App:HF_State}.

\subsection{Shallow circuit structures}
\label{subsec:shallow-circuit}

The importance of having very shallow parameterized quantum circuits for enhancing the performance of VQE algorithms has been extensively discussed in the literature~\cite{McClean_2016, McClean_barren, Zapata_circuits, HardEff}. In certain cases, it has even been the bottleneck for the implementation of VQE algorithms on current NISQ devices~\cite{NISQ_review, VQE_review}. For example, Kandala \textit{et al.}~\cite{HardEff} proposed the hardware-efficient ansatz which needed $504$ single-qubit and $106$ two-qubit gates on the statevector simulator to reach chemical accuracy for \ce{BeH2} using $6$ qubits. But for the hardware experiment they reduced the number of gates to $90$ single-qubit and $15$ two-qubit gates to obtain meaningful results on NISQ devices with error mitigation techniques~\cite{HardEff}.

\begin{figure}
\includegraphics[width=\columnwidth]{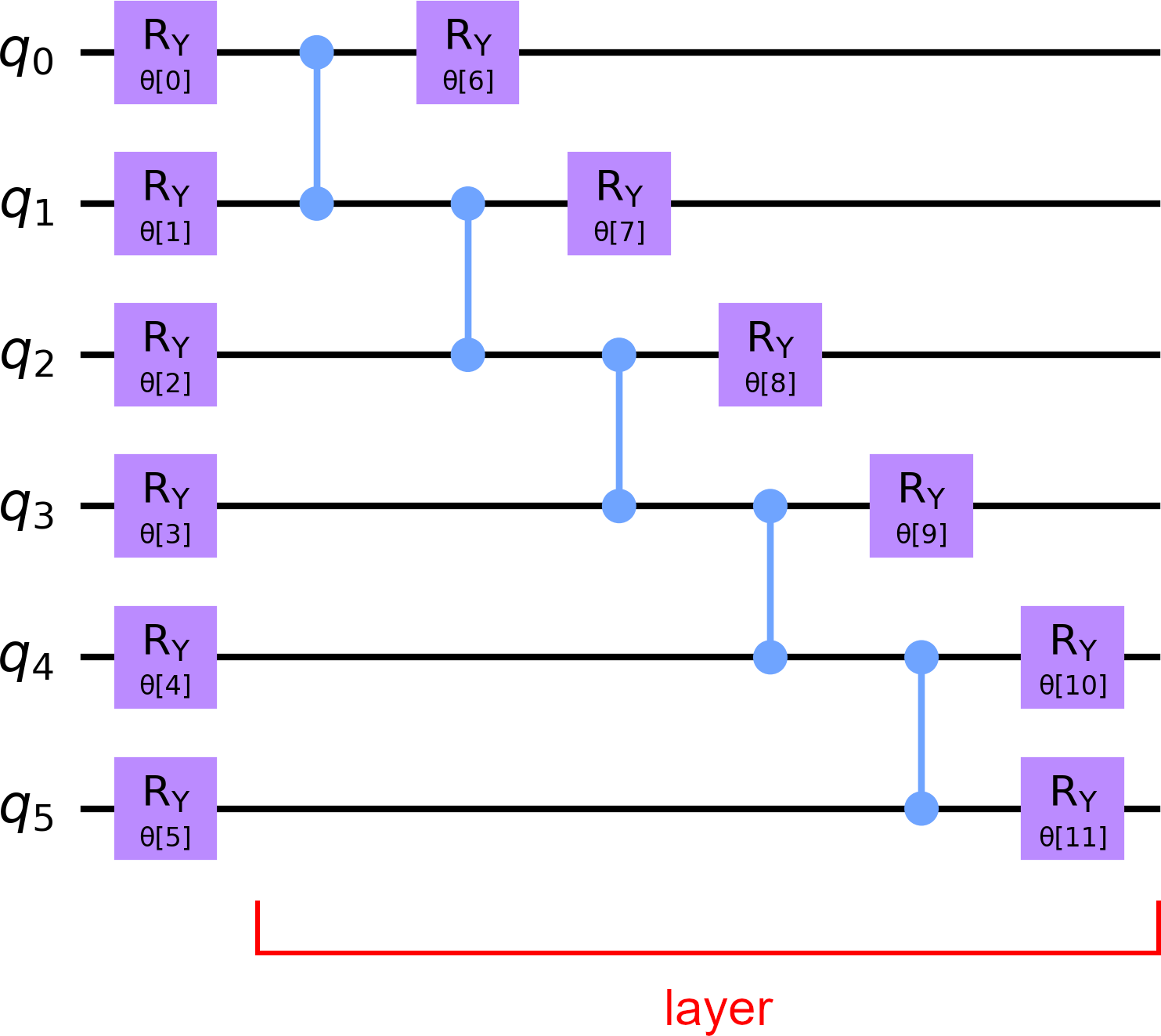}
\caption{The parameterized circuit ansatz used in this work and the definition of layer within this structure.}
\label{fig:6qubit_circuit}
\end{figure}

Different tools have been proposed to benchmark circuit architectures~\cite{Zapata_circuits, effect, ancilla_best}. An important characteristic of each circuit architecture is its entangling map that determines which qubits are entangled at each layer. The choice of the entangling map has also been studied~\cite{100qubit_VQE, Zapata_circuits} and incorporates a trade-off between the circuit depth and expressivity of the ansatz~\cite{100qubit_VQE, Zapata_circuits, too_much}. Here, we employ the linear entangling map (see Fig.~\ref{fig:6qubit_circuit}), which facilitates the generation of long-range entanglement and enhances the convergence of the VQE optimization~\cite{100qubit_VQE}.

Inspired by the hardware-efficient ansatz~\cite{HardEff}, we try to identify a circuit structure which is already hardware-friendly, i.e., has shallow circuit depth and a linear entangling map. Fig.~\ref{fig:6qubit_circuit} shows the circuit structure of one layer that we identify for the 6-qubit system as a good choice, i.e., that gives good fidelity values with shallow circuit depths. Here, we select CZ as the entangling gates, with the rationale explained in the next section. 

\begin{figure}
\includegraphics[width=\columnwidth]{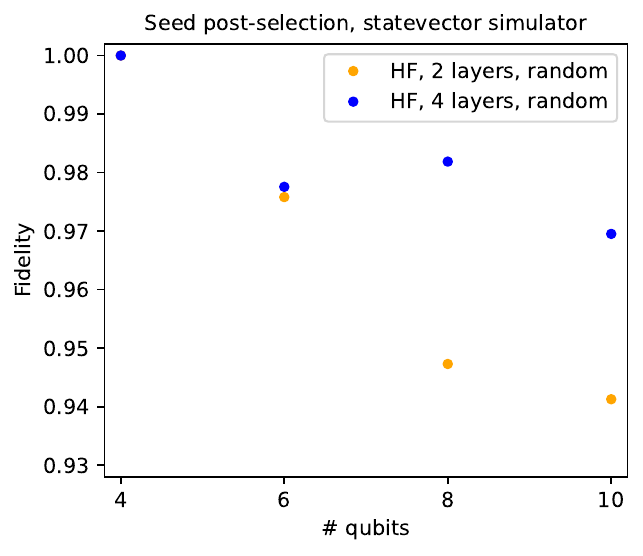}
\caption{The best achieved fidelities with post-selection over the seeds for our systems with 4, 6, 8, and 10 qubits on the statevector simulator.}
\label{fig:maxFid}
\end{figure}

This choice is beneficial for IBM's latest Heron processors, which implement CZ as a native gate. Remarkably, when we increase the system size to $8$ and $10$ qubits, we still get good fidelities above $0.94$ (see Fig.~\ref{fig:maxFid}) between the true (obtained from exact diagonalization) and VQE optimized ground state of our systems. For example, for the $10$-qubit system with $2$ and $4$ layers the fidelity is $0.941$ and $0.970$, respectively. Depending on the specific use case of the ground state approximation, one could decide whether it is beneficial to expand the circuit-depth by adding $2$ circuit layers for a $0.029$ increase in the fidelity~\cite{acc_critera}.

\begin{figure}[ht]
\includegraphics[width=\columnwidth]{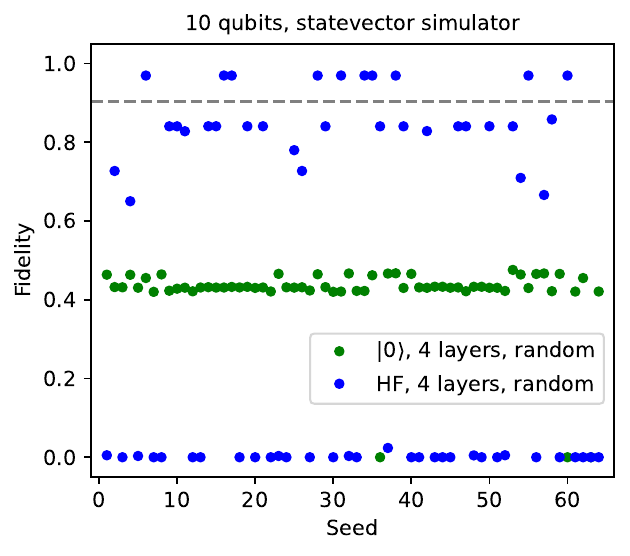}
\caption{The effect of the initialization of the parameterized set of gates (different seed choices) on the achieved fidelities with zero (green) or HF (blue) initial state for our system with 10 qubits on the statevector simulator. The dashed line represents the fidelity of the HF initial state.}
\label{fig:mf_2pi}
\end{figure}

For the case of $10$ qubits with $4$ layers, we have $50$ parameterized single-qubit gates that we need to optimize. Usually, the starting values of these gates are randomly chosen from the interval $[0, 2 \pi)$. Depending on the choice of the seed, a different set of the starting random values are applied to the parameterized gates which highly affects the performance of the algorithm and the fidelities that it achieves (blue dots in Fig.~\ref{fig:mf_2pi}). For many of the seeds, the optimization becomes trapped in local minima, leading to low fidelity values. This issue can be mitigated by increasing the number of parameters in the circuit, allowing the optimizer to explore more directions in the state space and escape spurious local minima, as demonstrated by Larocca \textit{et al.}~\cite{larocca-23-local-minima}. However, an increase in parameters gives rise to the issue of barren plateaus and may lead to a poor condition number for the Hessian matrix, thereby posing challenges for conventional gradient-based optimizers~\cite{LA_initial}.

To assess the impact of qubit initialization, we conducted additional simulations using the zero state as the initial state. However, we observed consistently low fidelity values for all seeds, represented by the green dots in Fig.~\ref{fig:mf_2pi}. In Fig.~\ref{fig:maxFid}, we have post-selected an optimal choice for the seed (i.e., $17$ in Fig.~\ref{fig:mf_2pi}) and the associated random starting values for the parameterized gates that could result in good fidelities after convergence of the optimization has been reached. The optimization settings are detailed in Appendix~\ref{App:NumSim}. This practice of post-selection has been employed in the literature~\cite{100qubit_VQE, Identity}. Given the inherent complexity of such post-selection, we explore methods to streamline this procedure, which is the focus of the next section.

\subsection{Initialization Strategy}
\label{subsec:IdBlock}

\begin{figure}[ht]
\includegraphics[width=\columnwidth]{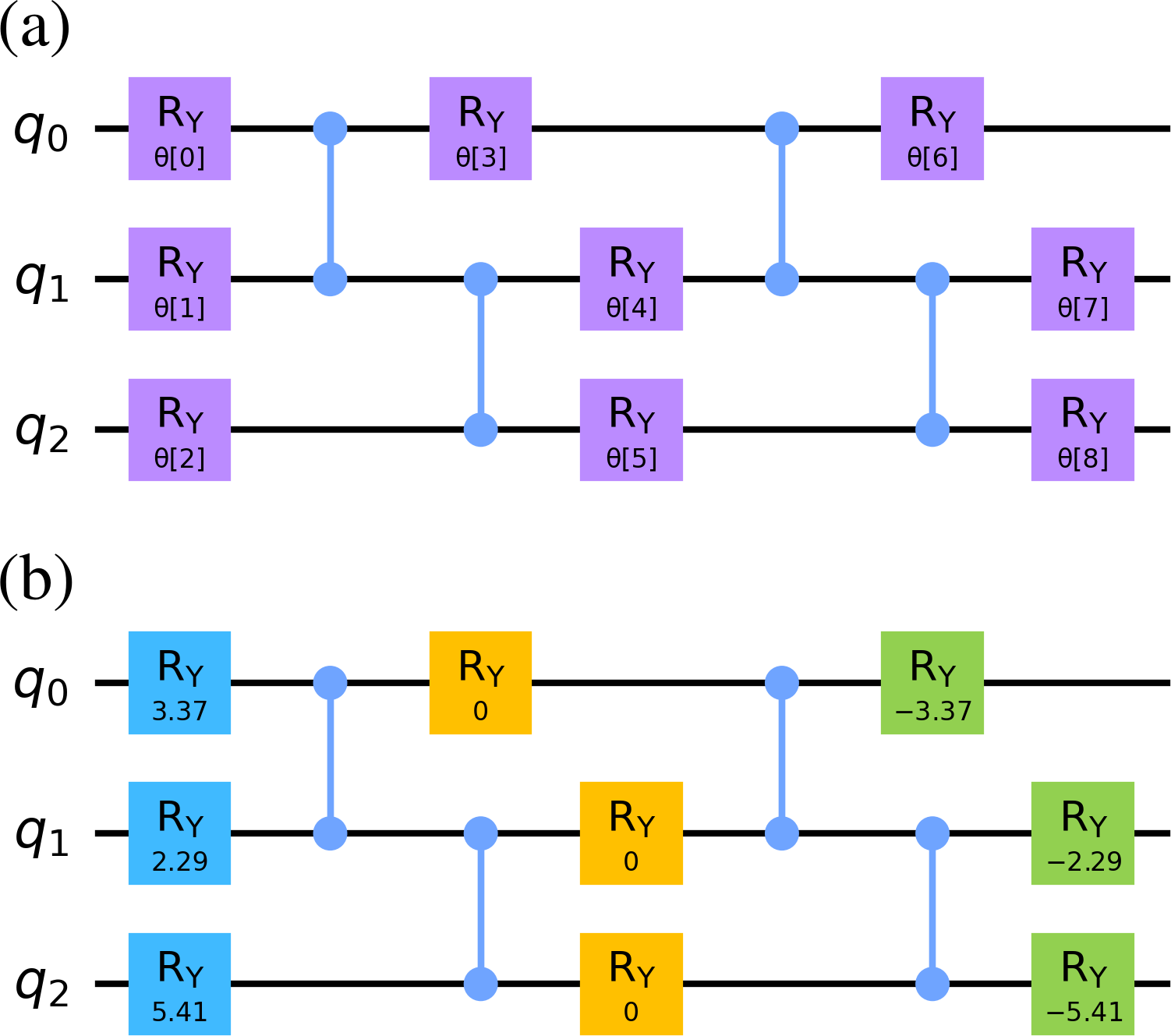}
\caption{(a) A 3-qubit circuit with two layers. (b) An example of the onion-initialization strategy where the parameters of the first (blue) and last (green) layers are set to random numbers ($\theta_0 = 3.37$, $\theta_1 = 2.29$, $\theta_2 = 5.41$) and their respective negative complements ($\theta_6 = -3.37$, $\theta_7 = -2.29$, $\theta_8 = -5.41$), while the intermediate layers (yellow) are initialized with zeros ($\theta_3 = 0$, $\theta_4 = 0$, $\theta_5 = 0$).}
\label{fig:IdBlock}
\end{figure}

\begin{figure}[ht]
\includegraphics[width=\columnwidth]{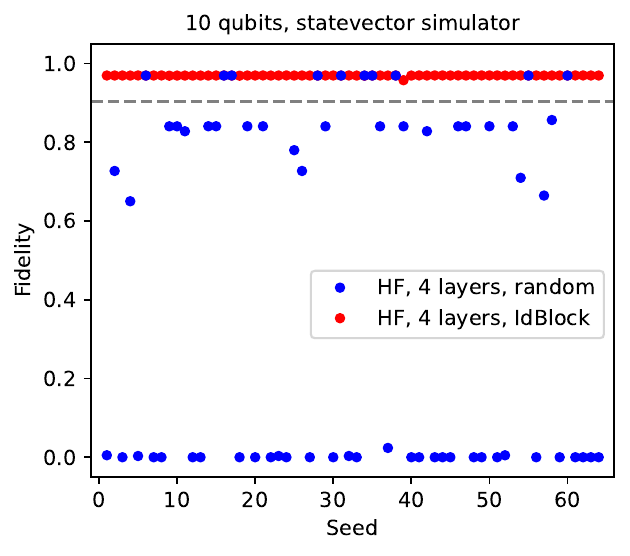}
\caption{The performance of the OnionVQE with the HF entangled initial state, the shallow ansatz with $4$ layers and the proposed onion-initialization of the parameterized gates (see Fig.~\ref{fig:IdBlock}) (red) versus the initialization of all parameterized gates to random values (blue) for our system with 10 qubits on the statevector simulator. The dashed line represents the fidelity of the HF initial state.}
\label{fig:VQE_opt_str}
\end{figure}

Many studies have identified the problem of how to choose the starting values of parameterized gates~\cite{McClean_2016, McClean_barren, unified, LA_initial} and have emphasized the importance of defining good initialization strategies. However, not many options have been proposed on how to do this. One approach is to initialize the parameterized gates close to identity~\cite{Identity}, and another is to apply the Identity Block initialization~\cite{Identity_Block_paper, wiersema2020identity} (here referred to as IdBlock). The idea of the IdBlock is to choose the parameterized gates in such a way that the whole unitary operation of the parameterized part of the circuit remains close to the identity operator. This procedure, according to the work of Grant \textit{et al.}~\cite{Identity_Block_paper}, helps keep gradients large at the first optimization step, leading to lower chances of encountering vanishing gradients and the associated barren plateaus during the subsequent optimization steps.

In the aforementioned study, the idea is to split the parameterized unitary $U$ into a product of two operators $U_1$ and $U_2$. Next, the parameterized gates are randomly initialized for $U_1$, and then, the parameterized gates in $U_2$ are selected in such a way that the whole operation of $U = U_2 U_1$ is close to identity. But given that the parameterized circuits typically have both single-qubit and two-qubit gates, finding $U_2$ is not an easy task in general. This makes the method a bit challenging to implement in practice. One special case of this approach that has been examined in the literature involves initializing all the parameterized gates to $\pi$~\cite{wiersema2020identity}.

In this paper, though, due to the simplicity of our circuit structures, we apply a simple trick. We set the parameters of the first and last layers to random numbers and to their respective negative complements. Simultaneously, we initialize the intermediate layers with zeros. This procedure yields the identity operator due to the cancellation of CZ entangling gates, which is the reason for choosing this type of entangling gates as mentioned in the previous section. To illustrate, let's assume that we have only a 3-qubit circuit with 2 layers as shown in Fig.~\ref{fig:IdBlock}. Following the IdBlock initialization strategy, we choose $U_1$ to be just the first application of $R_y$ gates, which means that $\theta_0, \theta_1$, and $\theta_2$ will be chosen randomly. Then, we set the last layer of Ry gates, namely $\theta_6, \theta_7$, and $\theta_8$ in Fig.~\ref{fig:IdBlock}, to $-\theta_0, -\theta_1$, and $-\theta_2$, respectively. Given that we have an even number of layers we set all the parameterized gates in between the first and last application of $R_y$ gates to exactly zero, and thus, the CZ entangling gates cancel out. As a result, the unitary operation $U$ of the parameterized part of the circuit is initialized to be the identity operator. We refer to this initialization strategy as $\textit{onion-initialization}.$

Remarkably, applying the onion-initialization strategy to the very shallow circuit architecture with the HF entangled state as the starting initial state, we are able to overcome the need of post-selection over the seeds. For the $10$-qubit system with $4$ layers, we consistently reach fidelity of $F = 0.969$ on the statevector simulator independent of the choice of the seed, as shown by red dots in Fig.~\ref{fig:VQE_opt_str}. This is contrary to the random initialization (blue dots in Fig.~\ref{fig:VQE_opt_str}) where only 16\% of the seeds exceed a fidelity of the HF initial state $F = 0.902$.

To quantify the relative importance of fidelity and entanglement properties of the initial state, we repeated the simulations depicted in Fig.~\ref{fig:VQE_opt_str} for Hamiltonians rotated to the MO basis, where the HF state transforms into a product state. However, we observe that the optimization halts after the first iteration due to vanishing gradients. This suggests that the entanglement of the initial state is crucial for achieving non-vanishing gradients in the initial iteration for our particular systems, as discussed earlier in Section~\ref{subsec:mf}. Further investigations are necessary to confirm whether non-vanishing gradients can be achieved in the MO basis for different parameterized circuit structures. Similarly, when starting from an entangled initial state with low fidelity ($F = 0.0004$), the optimization doesn't reach high fidelities ($F < 0.1$ for all seeds). This highlights the importance of both fidelity and entanglement properties of the initial state for the success of OnionVQE.

The ability of the OnionVQE optimization strategy to generate good fidelities, independent of the seed choice (red dots in Fig.~\ref{fig:VQE_opt_str}), is attributed to the combination of the three components: the entangled HF initial state, the shallow parameterized quantum circuit architectures, and the onion-initialization of the parameterized unitary to become the identity operator. Potentially the proposed optimization strategy could work with a larger number of layers which has been studied to some extent in the original research on the Identity Block initialization strategy~\cite{Identity_Block_paper}. In this paper, we focus on shallow parameterized quantum circuits with a small number of layers~\cite{McClean_barren, layer-wise}, as this approach generally facilitates optimization and makes it more realistic for implementation on NISQ devices.

\section{OnionVQE optimization on IBM's fake backends}
\label{sec:Results_ibmFakeDevices}

\begin{figure}[ht]
\includegraphics[width=\columnwidth]{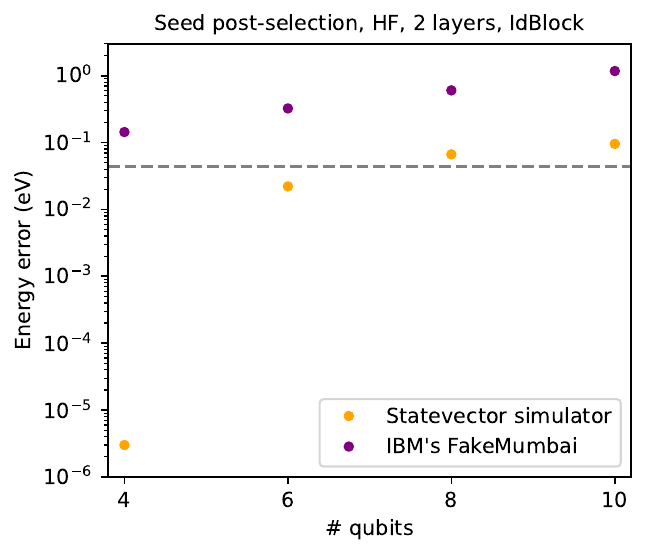}
\caption{The energy error of OnionVQE on the statevector simulator (orange) compared to IBM's FakeMumbai (purple) with post-selection over the seeds. The dashed line represents the chemical accuracy of $0.0016$ Ha.}
\label{fig:energy_error}
\end{figure}

\begin{figure}[ht]
\includegraphics[width=\columnwidth]{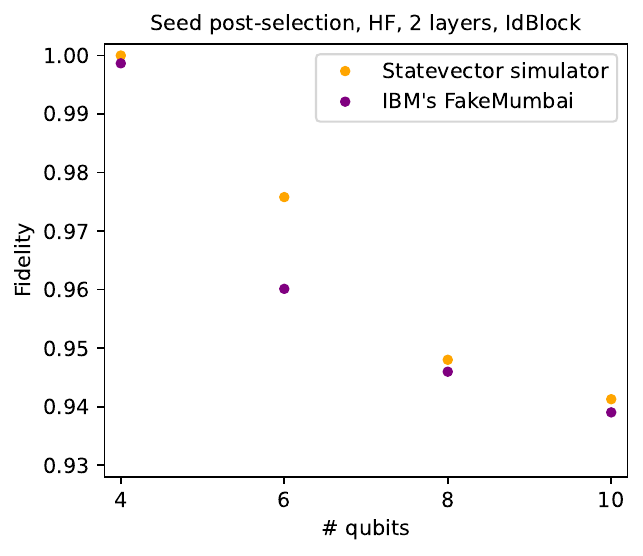}
\caption{The fidelities corresponding to the optimal values for parameterized gates obtained by running the OnionVQE optimization on the statevector simulator (orange) and on IBM's FakeMumbai (purple) with post-selection over the seeds. In both cases, the fidelities are computed by executing the circuit on the statevector simulator using the respective sets of optimal values for the parameterized gates.}
\label{fig:reps2_shots10240}
\end{figure}

In the previous section, we discussed the OnionVQE optimization strategy, which results in very shallow quantum circuits, making it promising for current noisy quantum devices. Here, we investigate the performance of the VQE optimization strategy with the entangled HF initial state, the shallow circuit structure, and the Identity Block initialization of the parameterized gates on IBM's fake backends.

IBM's fake backends mimic the real noise of IBM's quantum devices at a certain time using system snapshots. Such snapshots have been used for performing noisy simulations~\cite{OIST, H2_benchmark}, since they store important information such as the coupling map and qubit properties (T1, T2, gate and readout error probability of each qubit, and the gate length) of the quantum device~\cite{qiskit-tutorial-noise-simulation}. In this paper, we use the FakeMumbai backend for all our numerical simulations. Since in all cases the number of qubits in the device ($27$) is larger than the number of qubits needed for our systems ($4$---$10$), we can pre-select the qubits on the device used in the noisy simulations (see Fig.~\ref{Afig:fakeMumbai} in Appendix~\ref{App:FakeDevice}).

In Fig.~\ref{fig:energy_error}, we plot the energy error from running the OnionVQE optimization on IBM's FakeMumbai (purple) versus the optimization on the statevector simulator (orange), both with $2$ layers and post-selection over the seeds. For the optimization, we used the BFGS~\cite{L_BFGS_B} optimizer with the statevector simulator and the NFT~\cite{NFT} optimizer with IBM's FakeMumbai, respectively (refer to Appendix~\ref{App:NumSim} for additional details). The simulations on IBM's FakeMumbai were performed with 10240 shots and without employing error mitigation techniques. The energy error depicted in Fig.~\ref{fig:energy_error} increases to the order of $10^{-1}$~eV on IBM's FakeMumbai (purple) compared to $10^{-6}$~eV for the 4-qubit system and $10^{-2}$~eV for the 6, 8, and 10-qubit systems, respectively, on the statevector simulator (orange). Overall, we observe an approximately tenfold increase in the energy error on IBM's FakeMumbai compared to the noiseless statevector simulator.

To assess the similarity between the fidelity values obtained from executing OnionVQE on IBM's FakeMumbai and those from the statevector simulator, we utilize the optimal values for the parameterized gates obtained in both scenarios. Subsequently, in Fig.~\ref{fig:reps2_shots10240}, we plot the respective fidelities obtained by simulating the circuit on the statevector simulator with these two distinct sets of optimal values for the parameterized gates, applying post-selection over the seeds. We observe only a $\sim 1\%$ drop in fidelity values due to noise on IBM's FakeMumbai (purple) compared to the noiseless results from the statevector simulator (orange). For the 10-qubit system with 2 layers, the fraction of seeds exceeding a fidelity of the HF initial state ($F = 0.902$) is 44\% and 6\% for IdBlock and random initializations, respectively, compared to 100\% and 20\% achieved on the statevector simulator. This suggests that OnionVQE is somewhat resilient to noise and can achieve good fidelities on noisy devices, despite the fact that the algorithm only minimizes energy.

\section{Conclusions and Outlook}
\label{sec:Conclusions}

In this work, we propose a VQE optimization strategy that generates good ground state approximations (achieving fidelity values above $0.94$ for all the systems studied) with shallow circuit depths and a small number of gates. This makes it favorable for implementation on current NISQ devices. We applied the OnionVQE optimization strategy to IBM's FakeMumbai and observed an approximately tenfold increase in energy error when compared to the results from a noiseless statevector simulator (see Fig.~\ref{fig:energy_error}). Intriguingly, evaluating the circuit on a statevector simulator with the optimal values obtained from IBM's FakeMumbai results in only a $\sim 1\%$ drop in fidelities compared to the noiseless results with statevector-derived optimal values (see Fig.~\ref{fig:reps2_shots10240}). This suggests that even without any error mitigation techniques, the OnionVQE optimization strategy on the FakeMumbai backend results in a good set of values for the parameterized gates to match OnionVQE's performance on a noiseless statevector simulator.

The OnionVQE optimization strategy integrates various concepts proposed and thoroughly explored in the literature~\cite{LA_initial, Zapata_circuits, Identity_Block_paper}. Our results indicate that these concepts could help with certain issues commonly occurring in VQE algorithms. Specifically, the recent work of Leone \textit{et al.}~\cite{LA_initial} focuses on the effect of the choice of the initial state on VQE optimization. The Leone \textit{et al.} conclude that a non-trivial initial state, which generates entanglement and has some \textit{a priori} connection to the studied system, improves the performance. The importance of an entangled initial state has also been reported in other studies~\cite{Identity_Block_paper}.

Next, the importance of choosing a shallow circuit structure has extensively been discussed~\cite{McClean_2016, McClean_barren, HardEff, Zapata_circuits, Sim_2021, layer-wise}. All of the aforementioned studies showcase the importance of reducing the circuit depth and number of entangling gates. Finally, the choice of the starting values of the parameterized gates has also been highlighted, since it leads to one of the main challenges with VQE optimization, the barren plateau problem~\cite{McClean_barren, unified, LA_initial}. Grant \textit{et al.}~\cite{Identity_Block_paper} proposed the Identity Block initialization which helps obtain a non-vanishing gradient for most parameters at the first iteration of the VQE optimization.

The OnionVQE optimization strategy combines all three aforementioned mechanisms which results in a VQE algorithm with shallow circuit depth and small number of entangling gates that eliminates the need for post-selection of the starting values of the parameterized gates and the related problem with barren plateaus. Moreover, the proposed algorithm is independent of the choice of the Hamiltonian compared to other VQE algorithms in the literature that, e.g., exploit properties of the system under investigation~\cite{VQE_symmetries}. Finally, the OnionVQE optimization strategy yields shallow circuit structures suitable for current NISQ devices without any further adjustments. This contrasts with the work of Kandala \textit{et al.}~\cite{HardEff}, where the authors had to reduce both the number of entangling gates and the circuit depth to execute the VQE algorithm on noisy quantum devices.

Future research directions include exploring the performance improvement of OnionVQE with error mitigation techniques. Another avenue involves investigating optimal qubit pre-selection from various IBM devices and comparing fidelities with and without error mitigation. The OnionVQE algorithm achieves chemical accuracy of $0.0016$ Ha~\cite{HardEff} only for our $4$- and $6$-qubit systems on the statevector simulator. Further research is needed to extend this accuracy to larger system sizes.

Another crucial aspect is to explore the effects of choosing different initial states and initialization strategies. This involves using states other than HF and various entangled states. Additionally, the onion-initialization showed comparable performance to initializing all gates close to zero~\cite{Identity} for our particular systems. In contrast, initializing them close to $\pi$~\cite{wiersema2020identity} resulted in inferior performance for the 10-qubit system with 4 layers. In this case, the fraction of seeds exceeding a fidelity of the HF initial state ($F = 0.902$) decreased to 34\%, as opposed to the 100\% achieved with both the onion-initialization and the initialization close to zero. Further benchmarking would be necessary to compare these different approaches.

As already emphasized throughout this paper, the OnionVQE optimization strategy does not leverage any properties or symmetries of the studied system. It would be important to examine its performance for a broader range of systems, such as periodic solids and molecular Hamiltonians. Also, it would be interesting to explore how the algorithm behaves once we increase the system size beyond $10$ qubits. Another relevant topic to investigate is to better understand the effect of the decreasing HF fidelity values for larger systems~\cite{evidence_paper, Google_overlaps} on the performance of OnionVQE.

Finally, as discussed in more detail in Appendix~\ref{App:NumSim}, the choice of the optimizer seems to play an important role in the performance of OnionVQE. For example, the NFT optimizer seemed to perform better on IBM's FakeMumbai in comparison to other optimizers, such as SPSA~\cite{SPSA}. Further numerical benchmarking would be necessary to validate this observation.

In recent years, the utility of VQE algorithms has been thoroughly studied~\cite{acc_critera, Tilly_2022, LA_initial, purification}. A primary driver for this extensive examination is the challenges encountered in VQE optimization algorithms~\cite{McClean_2016, McClean_barren, unified, Gonthier_2022} and the necessity for error mitigation techniques~\cite{HardEff, H2_benchmark, HF_Google}. However, for quantum error correction to be realized, NISQ devices must consistently demonstrate reliable performance~\cite{preskill2018keynote}. This paper proposes the OnionVQE optimization strategy, which is capable of generating good ground state approximations (achieving fidelities above $0.94$ for all systems studied) with shallow circuit depths, even on current NISQ devices.

Our study underscores the significance of tailoring quantum algorithms to the limited capabilities of noisy quantum devices, with a particular focus on circuit depth. This customization approach is essential for improving result quality. To maximize the likelihood of success for NISQ and showcase the potential benefits of quantum computation, it is imperative to develop algorithms specifically suited to the current capabilities of quantum hardware.

\section{Acknowledgements}

This work was funded by Robert Bosch LLC, Research and Technology Center, where K.G. was a research intern. K.G. also acknowledges support from: European Union's Horizon 2020 research and innovation programme under the Marie Skłodowska-Curie grant agreement No. 847517, and ERC AdG NOQIA; Ministerio de Ciencia y Innovation Agencia Estatal de Investigaciones (PGC2018-097027-B-I00/10.13039/501100011033, CEX2019-000910-S/10.13039/501100011033, Plan National FIDEUA PID2019-106901GB-I00, FPI, QUANTERA MAQS PCI2019-111828-2, QUANTERA DYNAMITE PCI2022-132919,  Proyectos de I+D+I “Retos Colaboración” QUSPIN RTC2019-007196-7); MICIIN with funding from the European Union NextGenerationEU (PRTR-C17.I1) and by Generalitat de Catalunya;  Fundació Cellex; Fundació Mir-Puig; Generalitat de Catalunya (European Social Fund FEDER and CERCA program, AGAUR Grant No. 2021 SGR 01452, QuantumCAT \ U16-011424, co-funded by ERDF Operational Program of Catalonia 2014-2020); Barcelona Supercomputing Center MareNostrum (FI-2023-1-0013); EU (PASQuanS2.1, 101113690); EU Horizon 2020 FET-OPEN OPTOlogic (Grant No 899794); EU Horizon Europe Program (Grant Agreement 101080086 — NeQST), National Science Centre, Poland (Symfonia Grant No. 2016/20/W/ST4/00314); ICFO Internal “QuantumGaudi” project; European Union’s Horizon 2020 research and innovation program under the Marie-Skłodowska-Curie grant agreement No 101029393 (STREDCH) and No 847648  (“La Caixa” Junior Leaders fellowships ID100010434: LCF/BQ/PI19/11690013, LCF/BQ/PI20/11760031,  LCF/BQ/PR20/11770012, LCF/BQ/PR21/11840013). Views and opinions expressed are, however, those of the authors only and do not necessarily reflect those of the European Union, European Commission, European Climate, Infrastructure and Environment Executive Agency (CINEA), nor any other granting authority.  Neither the European Union nor any granting authority can be held responsible for them.

J.S., M.A. and T.E. gratefully acknowledge support from the German Federal Ministry of Education and Research (BMBF) under project No. 13N15574. We thank Sophie Beck and Alexander Hampel for fruitful expert discussions.

\renewcommand{\bibsection}{\section*{References}}

%\bibliography{main.bib}

\clearpage

\appendix

\section{HF state}
\label{App:HF_State}

\begin{figure}[ht]
\includegraphics[width=\columnwidth]{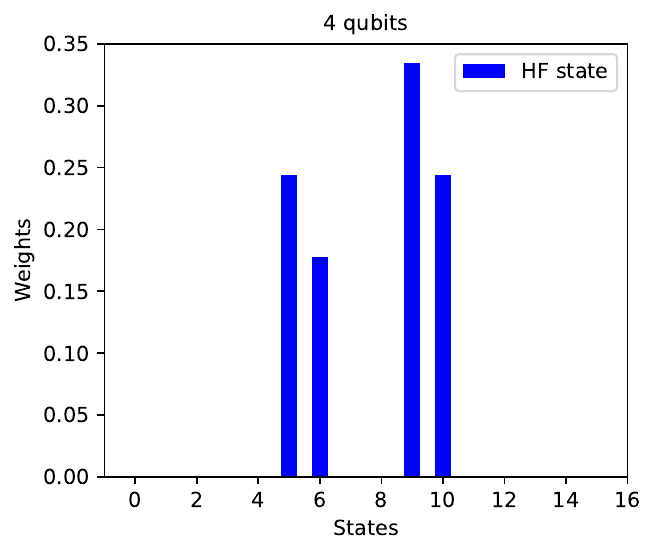}
\caption{The statevector of the Hartree-Fock (HF) state, used as the initial state for our system with $4$ qubits.}
\label{Afig:mf_4qubit}
\end{figure}

\begin{figure}[ht]
\includegraphics[width=\columnwidth]{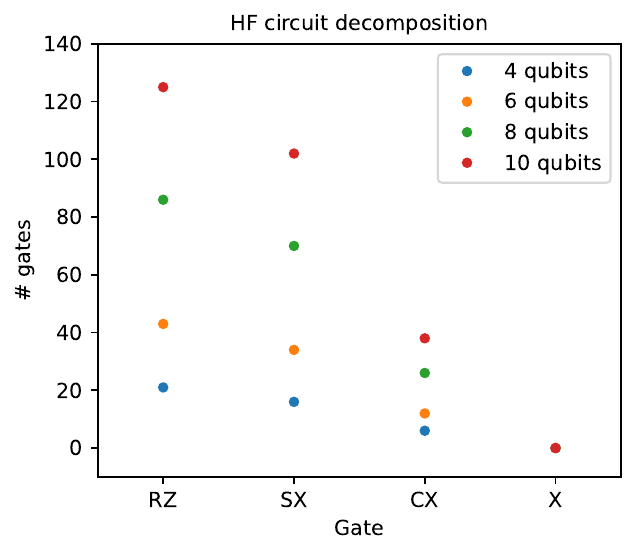}
\caption{Gate counts in the decomposed Hartree-Fock (HF) circuits, serving as initial states for our systems with 4, 6, 8, and 10 qubits.}
\label{Afig:mf_decomposition}
\end{figure}

In Fig.~\ref{Afig:mf_4qubit}, we plot the statevector of the HF state used as initial state for the 4-qubit circuit. In Fig.~\ref{Afig:mf_decomposition}, we show the HF decomposition in terms of number of gates that compile the HF quantum circuit in Qiskit for all 4, 6, 8, and 10-qubit circuits used.

\section{IBM's FakeMumbai}
\label{App:FakeDevice}

\begin{figure}[ht]
\includegraphics[width=\columnwidth]{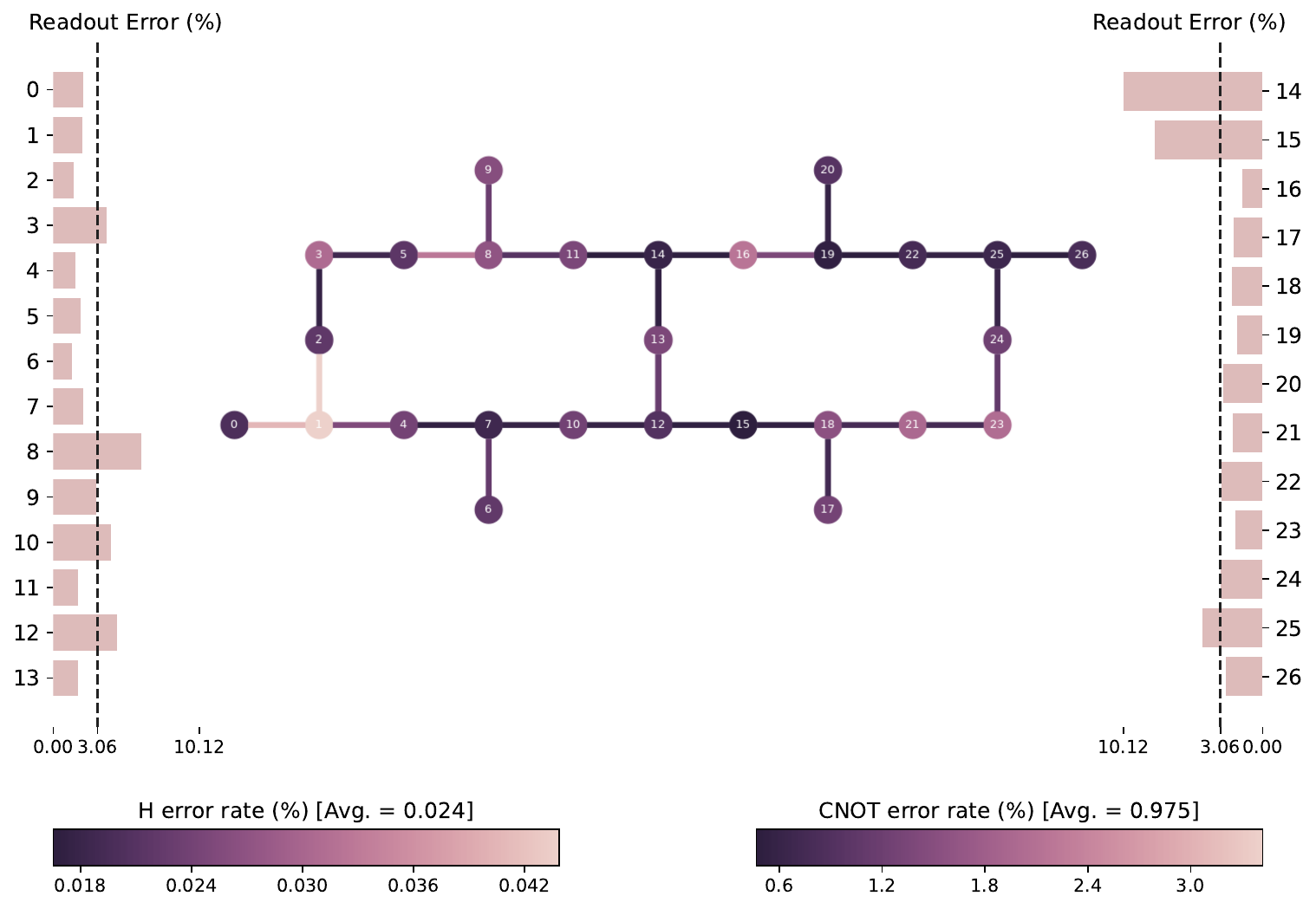}
\caption{Error map of IBM's FakeMumbai device.}
\label{Afig:fakeMumbai}
\end{figure}

In Fig.~\ref{Afig:fakeMumbai}, we present the error map of IBM's FakeMumbai device. For the noisy simulations on IBM's FakeMumbai, we pre-selected the qubit layouts. Specifically, for the 4, 6, 8, and 10-qubit circuits, we utilized the layouts $[4,7,10,12]$, $[4,7,10,12,13,14]$, $[1,4,7,10,12,13,14,16]$, and $[3,2,1,4,7,10,12,13,14,16]$, respectively. We highlight the existence of various suitable layouts, as we randomly selected one from the many possible configurations.

Apart from the read-out, single-qubit and two-qubit gate errors reported in Fig.~\ref{Afig:fakeMumbai}, the decoherence times are also important for the implementations. The FakeMumbai used in this work with the layout $[3,2,1,4,7,10,12,13,14,16]$ has energy relaxation time $T_1=120 \pm 28$ $\mu$s and dephasing time $T_2=124 \pm 50$ $\mu$s.

For the noisy simulation, we used $10240$ shots. If we use only $1024$ shots and once again explore the performance of the optimal values of the parameterized gates with the statevector simulator, we see that the fidelity drops to $0.949$ and $0.897$ from $0.999$ and $0.939$ for the $4$ and $10$-qubit circuits, respectively. For the $10$-qubit circuit with $10240$ shots, we get approximately the same results for $2$ and $4$ layers, namely $0.939$ and $0.936$, respectively.

\section{Hamiltonian construction and solving the Anderson impurity model}
\label{App:Hamiltonian}

The model Hamiltonian was derived from the cuprate material \ce{CaCuO2}, which crystallizes in a tetragonal structure with $P4/mmm$-symmetry. Square planar sheets of \ce{CuO2} are stacked along the $z$ direction, interspersed with Ca cations~\cite{karpinski_single_1994}. A single electronic band across the Fermi level arising from the Cu $d_{x^2-y^2}$ state is commonly believed to be responsible for correlation effect resulting in the intricate physics of \ce{CaCuO2} and similar cuprate (superconducting) materials~\cite{karp_many-body_2020, karp_superconductivity_2022}.

We use density functional theory (DFT) calculations to construct the effective Hamiltonian. The Quantum ESPRESSO plane wave package~\cite{giannozzi_quantum_2009} together with the Perdew-Burke-Ernzerhof approximation to the exchange-correlation functional~\cite{perdew_generalized_1996} and norm-conserving pseudopotentials~\cite{van_setten_pseudodojo_2018} are used for the DFT calculations. We employ a plane-wave cutoff energy of 100~Ry for the wave function in conjunction with a $k$-point mesh of $12\times 12\times 12$ to obtain converged results. A single-orbital tight-binding model is constructed by Wannierizing the $d_{x^2-y^2}$ band crossing the Fermi level using the Wannier90 package~\cite{pizzi_wannier90_2020}. We convert the orbital with wan2respack~\cite{kurita_interface_2023} to serve as input for RESPACK~\cite{nakamura-20-respack} and obtain the screened Coulomb interaction parameters based on the constrained random phase approximation (cRPA). We include 80 virtual orbitals in addition to the 20 fully occupied states and employ a plane-wave cutoff energy of 10~Ry for the polarizability, resulting in well converged screened interaction parameters. The static interaction parameter $U$ in our Hamiltonian is obtained by taking the zero-frequency limit of the real part of the frequency dependent onsite interaction $U=\lim \limits_{\omega \to 0} \Re[U_\text{cRPA}(\omega)]$.

The Hubbard Hamiltonian is mapped to an Anderson impurity model with open boundary conditions and solved with dynamical mean field theory (DMFT) as implemented in solid\_dmft as part of the TRIQS package~\cite{Merkel2022}, using the associated continuous-time quantum Monte Carlo (QMC) impurity solver based on hybridization expansion (CTHYB)~\cite{Seth2016274} at an inverse temperature of $\beta=30/\text{eV}$ (around 387~K). We perform 15 DMFT cycles with \num{100e6} QMC samples each, leading to a sufficiently converged impurity Green's function. The converged CTHYB results are then used as a starting point for an exact diagonalization calculation at zero temperature.

We perform 10 DMFT cycles using an exact diagonalization impurity solver, which calculates the statevector of the ground and excited states by diagonalizing the Hamiltonian of the Anderson impurity model, and finally computes the Green's function in form of a Lehmann representation. The DMFT cycle is performed in imaginary frequency, where the grid points are given by the Matsubara frequencies of the CTHYB-DMFT calculation at $\beta=30/\text{eV}$. The parameters of the Anderson impurity models used for the exact diagonalization solver and the VQE calculations in this study are obtained by performing a fitting procedure to the hybridization function given in the DMFT cycle. The intricate details of this procedure are described elsewhere~\cite{selisko-24-dmft}. The Anderson impurity model is chosen to have chain topology and is given by:
\begin{align}
    \hat{H} = & \sum_\sigma\sum_{i=1}^{n_b}\left[\epsilon_{i} n_{i\sigma} + V_i(c^\dagger_{i\sigma} c_{(i-1)\sigma} + c^\dagger_{(i-1)\sigma} c_{i\sigma})\right]  \nonumber\\
    &+ U(n_{0\uparrow}n_{0\downarrow}) + \epsilon_{0} (n_{0\uparrow} + n_{0\downarrow}),
\end{align}
which results in an appropriate Hamiltonian for quantum computing calculations. The index $i$ refers to the bath $i$ with the site energies $\epsilon_{i}$, while index $0$ denotes the impurity site with $\epsilon_0$ being the impurity atomic level. $V_i$ denotes the hopping along the chain, while $U$ is the Hubbard interaction parameter obtained from cRPA.

In Table~\ref{table:Hamiltonians} we present the site energies $\epsilon_i$ and hopping along the chain $V_i$ for 4, 6, 8, and 10-qubit systems. The Hubbard interaction $U$ remains constant and equal to 3.11~eV for all the system sizes discussed here. The numbers of electrons in the ground state obtained by exact diagonalization are 2, 3, 4, and 4 for the 4, 6, 8, and 10-qubit systems, respectively.

\begin{table*}[ht]
    \centering
    \caption{The impurity atomic level $\epsilon_0$, the bath $i$ site energies $\epsilon_i$, and hopping along the chain $V_i$ for 4, 6, 8, and 10-qubit systems with 1, 2, 3, and 4 bath sites $n_b$, respectively. All values are in eV.}
    \label{table:Hamiltonians}
    \begingroup
    \setlength{\tabcolsep}{4pt}
    \renewcommand{\arraystretch}{1.25}
    \begin{tabular}{lrrrrrrrrrr}
        \hline
        $n_b$ & $\epsilon_0$ & $\epsilon_1$ & $\epsilon_2$ & $\epsilon_3$ & $\epsilon_4$ & $V_1$ & $V_2$ & $V_3$ & $V_4$ \\
        \hline
        \hline
        $1$ & $-1.201$ & $ 0.026$ & $      $ & $      $ & $      $ & $ 0.270$ & $      $ & $      $ & $      $ \\
        $2$ & $-1.312$ & $ 0.082$ & $-0.006$ & $      $ & $      $ & $ 0.513$ & $ 0.156$ & $      $ & $      $ \\
        $3$ & $-1.317$ & $ 0.063$ & $-0.109$ & $-0.009$ & $      $ & $ 0.739$ & $ 0.493$ & $ 0.178$ & $      $ \\
        $4$ & $-1.123$ & $ 0.026$ & $-0.171$ & $ 0.041$ & $ 0.016$ & $ 0.917$ & $ 1.011$ & $ 0.512$ & $ 0.139$ \\
        \hline
    \end{tabular}
    \endgroup
\end{table*}

\section{Qubit Hamiltonians}
\label{App:QHamiltonian}

\begin{figure}[ht]
\includegraphics[width=\columnwidth]{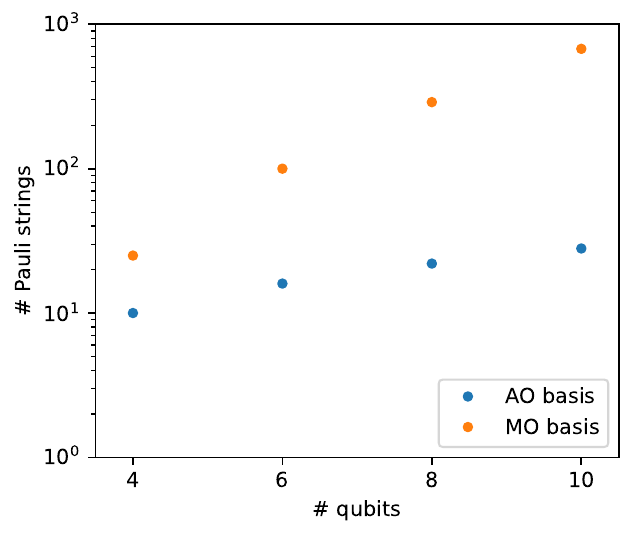}
\caption{Total number of Pauli strings in the atomic orbital (AO) (blue) and molecular orbital (MO) (orange) bases for our systems with 4, 6, 8, and 10 qubits.}
\label{Afig:Pauli_strings}
\end{figure}

\begin{figure}[ht]
\includegraphics[width=\columnwidth]{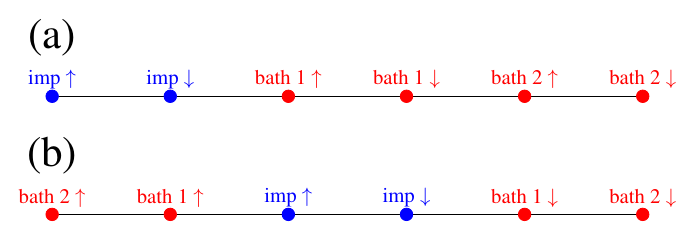}
\caption{(a) The Jordan-Wigner mapping of spin-orbitals to qubits for our system with $6$ qubits ($1$ impurity and $2$ bath sites). (b) The same mapping utilizing the re-numbering of the qubits.}
\label{Afig:qubit_ordering}
\end{figure}

Next, we use the Jordan-Wigner transformation in Qiskit~\cite{qiskit2024} to map the fermionic to qubit Hamiltonians, resulting in a relatively small number of Pauli strings (see Fig.~\ref{Afig:Pauli_strings}). The Jordan-Wigner transformation maps the spin-orbitals to qubits. For instance, it assigns the spin-up and spin-down orbitals of the impurity to qubits 0 and 1, and those of the two bath sites to qubits 2, 3, 4, and 5 (see Fig.~\ref{Afig:qubit_ordering}(a)).

Here, we re-number the qubits, starting with the chain of up-spin bath qubits in reversed order, followed by the impurity qubits (first up-spin, then down-spin), and finally the down-spin bath qubits (see Fig.~\ref{Afig:qubit_ordering}(b)). This reordering ensures that all terms in the Hamiltonian transform into nearest neighbor interactions along the chain. As a result, the Hamiltonian can be measured in three steps independent of the model's size, achieved by measuring all qubits once in the X, Y, and Z basis.

\section{Numerical simulations}
\label{App:NumSim}

For convenience we summarize some details of the numerical calculations already discussed in the main text. Throughout this paper, we employ the generalized Hartree-Fock form within the framework of the Hartree-Fock theory, which is computed with \textsc{pyscf}~\cite{pyscf}, resulting in fidelity values of $0.787$, $0.798$, $0.889$, and $0.902$ for the 4, 6, 8, and 10-qubit systems, respectively. For all simulations in this work, we used Qiskit~\cite{qiskit2024}. For the statevector simulator, the limited memory Broyden-Fletcher-Goldfarb-Shanno (BFGS) bound-constrained optimizer~\cite{L_BFGS_B} was employed, utilizing a finite difference gradient with a step size of $\epsilon=0.01$. Conversely, for simulations on IBM's FakeMumbai, the Nakanishi-Fujii-Todo (NFT) optimizer~\cite{NFT} was selected. We employed Qiskit's default convergence criteria for both optimizers, including a relative tolerance for termination set at $2.22 \times 10^{-15}$, and the maximum numbers of function evaluations and iterations both configured to 15000 (BFGS) and 1024 (NFT). At this point we would like to emphasize that NFT seemed to play an important role in achieving good fidelities on IBM's FakeMumbai in comparison to other optimizers, such as the Simultaneous Perturbation Stochastic Approximation (SPSA) optimizer~\cite{SPSA}. Further numerical benchmarking would be necessary to validate this observation.

Throughout this paper, we define the energy error as the difference between the computed VQE energy $E$ and the true ground state energy $E_0$. From the numerical simulations we performed, we observed that if we use the zero state as initial state with the proposed onion-initialization, we get a broad variation in fidelity values depending on the random seed. This behavior is similar to the effect of the random initialization of the parameterized gates illustrated in Fig.~\ref{fig:mf_2pi}. But interestingly, if we use the zero initial state with the CX linear entangling map instead of the CZ in Fig.~\ref{fig:6qubit_circuit} and with $7$ layers, we obtain $F = 0.9920$ and $E - E_0 = 0.00844$~eV for the $10$-qubit system using seed post-selection. These results are well within the chemical accuracy of $0.0016$~Ha.

\end{document}